\documentclass[10pt]{iopart}

\usepackage{bm} 
\usepackage{xcolor}
\usepackage{graphicx,caption}
\usepackage{epstopdf}
\usepackage[colorlinks,linkcolor=blue,anchorcolor=blue,citecolor=blue,urlcolor=blue]{hyperref}

\newcommand{\nn}{{\nonumber}}
\newcommand{\bea}{\begin{eqnarray}}
\newcommand{\eea}{\end{eqnarray}}
\newcommand{\ie}{\textit{i.e.{ }}}
\newcommand{\eg}{\textit{e.g.{ }}}

\newcommand{\up}{\uparrow}
\newcommand{\dn}{\downarrow}

\newcommand{\av}[1]{\langle #1 \rangle}

\begin{document}

\title{Anomalous isotope effect in BCS superconductors with two boson modes}
\author{Gan Sun$^1$, Pan-Xiao Lou$^1$, Sheng-Qiang Lai$^1$, Da Wang$^{1,2}$ and Qiang-Hua Wang$^{1,2}$}
\address{$^1$National Laboratory of Solid State Microstructures $\&$ School of Physics, Nanjing University, Nanjing 210093, China}
\address{$^2$Collaborative Innovation Center of Advanced Microstructures, Nanjing University, Nanjing 210093, China}
\ead{dawang@nju.edu.cn}
\ead{qhwang@nju.edu.cn}
\begin{abstract}
The isotope effect in the superconducting transition temperature is anomalous if the isotope coefficient $\alpha<0$ or $\alpha>1/2$.
In this work, we show that such anomalous behaviors can naturally arise within the Bardeen-Cooper-Schrieffer framework if both phonon and non-phonon modes coexist.
Different from the case of the standard Eliashberg theory (with only phonon) in which $\alpha\le1/2$, the isotope coefficient can now take arbitrary values in the simultaneous presence of phonon and the other non-phonon mode.
In particular, most strikingly, a pair-breaking phonon can give rise to large isotope coefficient $\alpha>1/2$ if the {unconventional} superconductivity is mediated by the lower frequency non-phonon boson mode.
Based on our studies, implications on several families of superconductors are discussed.
\end{abstract}
\submitto{\NJP}
\maketitle

\section{Introduction}
Isotope effect \cite{Maxwell1950,Reynolds1950} is a cornerstone of the Bardeen-Cooper-Schrieffer (BCS) theory \cite{Bardeen1957} for phonon mediated superconductors.
It describes the change of transition temperature $T_c$ caused by isotope substitution.
The standard BCS theory gives $T_c = 1.13\Omega \mathrm{e}^{-1/\lambda}$ and thus predicts the isotope coefficient
\begin{eqnarray}
\alpha=-\frac{\partial \ln T_c}{\partial \ln M} =\frac12\frac{\partial\ln T_c}{\partial\ln \Omega}=\frac12,
\end{eqnarray}
where $M$ is the ion mass, $\Omega$ is the Debye frequency and $\lambda$ is the dimensionless isotope independent electron-phonon coupling constant.
Including the Coulomb pseudopotential \cite{Morel1962} $\mu^*=\mu/[1+\mu\ln(E_F/\Omega_D)]$ ($\mu$ is  defined at the Fermi energy $E_F$) will reduce the isotope coefficient even to negative values, as clarified in the more elaborated Eliashberg theory. \cite{Migdal1958,Eliashberg1960,Garland1963,Carbotte1990}

In real materials, the anomalous isotope effect (defined as $\alpha<0$ or $\alpha>1/2$) has been observed in many experiments.
(Although the negative isotope coefficient $\alpha<0$ is a standard behavior predicted by the Eliashberg theory, it can only occur at very low $T_c$ and thus is inconsistent with the experiments. Therefore, people still prefer to call $\alpha<0$ anomalous.)
Among all these materials, cuprates may be the most systematically studied in the past thirty years. \cite{SchriefferBook}
In cuprates, the isotope coefficient of the O-atoms in the CuO$_2$-plane \cite{Zech1994} is found to nearly vanish at optimal doping \cite{Batlogg1987} and to increase with decreasing $T_c$ either upon under- or over-doping, to values even larger than $1/2$. \cite{Keller2005,Keller2008}
Such an interesting observation has stimulated many theoretical works on the role of phonons in the superconductivity mechanism of cuprates, \cite{Tsuei1990,Carbotte1991,Schuttler1995,Bulut1996,Xing1999,Pringle2000,Devereaux2004,Dolgov2005,Newns2007,Honerkamp2007,Keller2008,Harshman2008,Johnston2010,Alexandrov2012,Muller2014} although spin and other types of fluctuations are also widely observed to be closely related to superconductivity. \cite{Dai1999,Tranquada2007,Carbotte2011,Fujita2012}
In fact, cuprates are not the only material having $\alpha>1/2$.
Such an anomalous behavior has also been observed in some iron- \cite{Khasanov2010} and C$_{60}$-based superconductors \cite{Gunnarsson1997,GinsbergBook,Ricco2008}.
In Sr$_2$RuO$_4$, even a similar scaling behavior of $\alpha$ versus $T_c$ was reported and drops to negative values near the maximum $T_c$. \cite{Mao2001}
Such a negative isotope coefficient cannot be explained by the standard Eliashberg theory where $\alpha$ can drop to negative values but only as $T_c$ diminishes. \cite{Carbotte1990}
Similarly, the inverse isotope effect has also been observed in iron-based superconductors \cite{Shirage2009} and PdH \cite{Skoskiewicz1972}.
In the latter case, the anomalous isotope effect has been attributed to anharmonic phonon effect. \cite{Matsushita1980,Klein1992,Errea2013}

How to understand these anomalous isotope effect in a unified way is a longstanding problem.
This issue has been explored rather extensively in the literature.
Among these studies, for cuprates in particular, some material dependent properties such as Van Hove singularity \cite{Tsuei1990,Xing1999}, pseudogap \cite{Pringle2000}, anharmonic phonon effect \cite{Schuttler1995,Newns2007}, or bipolaron \cite{Keller2008,Alexandrov2012,Muller2014} are proposed to be responsible for the anomalous isotope effect.
A somewhat more ``universal'' approach was to consider a pair-breaking non-phonon mode which is found to cause large $\alpha>1/2$. \cite{Carbotte1991,Dolgov2005,Harshman2008}
However, most of these studies assumed phonon mediated superconductivity with $\lambda_{\rm ph}>0$ (henceforth the subscript ph/nph stands for phonon/non-phonon in this work).
But for unconventional superconductivity (other than uniform s-wave pairing), $\lambda_{\rm ph}$ can be negative for a given phonon mode. \cite{Scalapino1986,Johnston2010}
See the appendix for several typical phonon modes in cuprates for example.
Another aspect for some of these works may be the heavy dependence on the material details, \eg complex electron band structures or electron-boson coupling functions $\alpha^2F$, to solve the Eliashberg equations.
But a universal understanding of the anomalous isotope effect is still lacking.

Motivated by these experimental and theoretical progresses, we ask a somewhat very simple question: {\it what happens in the BCS framework {(not only with s-wave pairing)} including two kinds of single-frequency bosons $\Omega_{1,2}$ with the electron-boson couplings $\lambda_{1,2}$?}
Thanks to the small number of model parameters, a thorough study becomes possible.
Interestingly, all types of values of $\alpha$ can appear within this simple approach.
In particular, we find a pair-breaking phonon can give rise to $\alpha>1/2$ as long as the other non-phonon mode has a lower frequency.
Implications on different superconducting materials will be discussed.

\section{BCS theory with two boson modes}
Boson mediated interactions are retarded.
One key point of the standard BCS theory\cite{Bardeen1957} is to simplify the frequency dependency of the boson mediated pairing interactions into the momentum space within an energy shell: $V_{\0k\0k'}=V\Theta(\Omega_D-|\varepsilon_{\0k}|) \Theta(\Omega_D-|\varepsilon_{\0k'}|)$ with $\Theta$ the step function {and $\varepsilon_{\0k}$ the normal state band dispersion}.
Following this idea directly, including two boson modes, we have the interaction Hamiltonian $H_I=-\sum_{\0k\0k'}V_{\0k\0k'}\Delta_{\0k}^\dag\Delta_{\0k'}$ with $\Delta_{\0k}$ the pairing operator and $V_{\0k\0k'}$ the pairing interaction given by
\begin{eqnarray}
V_{\0k\0k'}=&\sum_{m} f_m(\0k) f_m(\0k')\left[V_{m1} \Theta(\Omega_1-|\varepsilon_{\0k}|) \Theta(\Omega_1-|\varepsilon_{\0k'}|) \right. \nn\\
&\left.+V_{m2} \Theta(\Omega_2-|\varepsilon_{\0k}|) \Theta(\Omega_2-|\varepsilon_{\0k'}|)-U_{m}
\right],
\label{eq:model}
\end{eqnarray}
where the pairing interactions have been decoupled into different symmetry channels {(e.g. s-, p- or d-wave etc.)} labeled by the subscript $m$ with $f_m(\0k)$ defined as the form factor (the $m$-th eigenvector of $V_{\0k\0k'}$). $V_{m1}$ and $V_{m2}$ are effective interactions {(positive for attractive ones)} mediated by the $\Omega_1$ and $\Omega_2$ modes, respectively.
$U_m$ is the instantaneous interaction.
Without lost of generality, $\Omega_1\leq\Omega_2$ is always assumed in the following.

At $T=T_c$, we have the linearized gap equation $\Delta(\0k)=\sum_{\0k'}K(\0k,\0k')\Delta(\0k')$ where
\begin{eqnarray}
K(\0k,\0k')=\frac{V_{\0k\0k'}}{N_\0k}\frac{\tanh(\varepsilon_{\0k'}/2T)}{2\varepsilon_{\0k'}}.
\end{eqnarray}
where $N_\0k$ is the number of momentum $\0k$.
As a result of the three-piecewise behavior of $V_{\0k\0k'}$ in Eq.~\ref{eq:model}, $\Delta(\0k)$ can be approximated as a three-piecewise function.
Then, the momentum summation (replaced by energy integral with constant density of states) can be performed in three regimes: $(0,\Omega_1)$, $(\Omega_1,\Omega_2)$ and $(\Omega_2,E_F)$, respectively.
For the $m$-wave pairing (it means the pairing function is $f_m$), the kernel becomes (assuming $T_c\ll \Omega_1$)
\begin{eqnarray}
K=\left[\begin{array}{ccc}
(\lambda_1+\lambda_2-\mu)\ln\left(\frac{\kappa\Omega_1}{T}\right) & (\lambda_2-\mu)\ln\left(\frac{\Omega_2}{\Omega_1}\right) & -\mu\ln\left(\frac{E_F}{\Omega_2}\right) \\
(\lambda_2-\mu)\ln\left(\frac{\kappa\Omega_1}{T}\right) & (\lambda_2-\mu)\ln\left(\frac{\Omega_2}{\Omega_1}\right) & -\mu\ln\left(\frac{E_F}{\Omega_2}\right) \\
-\mu\ln\left(\frac{\kappa\Omega_1}{T}\right) & -\mu\ln\left(\frac{\Omega_2}{\Omega_1}\right) & -\mu\ln\left(\frac{E_F}{\Omega_2}\right) \\
\end{array}\right],
\end{eqnarray}
where $\kappa=2e^\gamma/\pi\approx1.13$, $\lambda_{1,2}=\+NV_{m1,2}\av{f_m^2}$, and $\mu=\+NU_m\av{f_m^2}$ with $\+N$ the electron density of states and the average is performed on the Fermi surface.
$T_c$ is then determined by letting the largest eigenvalue of $K$ to be $1$, \ie $\det(K-I)=0$, giving rise to the $T_c$-formula:
\begin{eqnarray} \label{eq:Tc}
T_c=\kappa\Omega_1\exp\left\{ - \frac{1+(\mu^*-\lambda_2)\ln\left(\frac{\Omega_2}{\Omega_1}\right)}{\lambda_1+\lambda_2-\mu^*+\lambda_1(\mu^*-\lambda_2)\ln\left(\frac{\Omega_2}{\Omega_1}\right)} \right\},
\end{eqnarray}
where $\mu^*$ is the Coulomb pseudopotential defined at $\Omega_2$ as usual, \ie $\mu^*=\mu/[1+\mu\ln(E_F/\Omega_2)]$. \cite{Morel1962}
It can be easily checked by setting $\lambda_1=0$ or $\lambda_2=0$, Eq.~\ref{eq:Tc} does reduce to the standard BCS result.
In fact, Eq.~\ref{eq:Tc} can be rewritten in a more familiar way: $T_c=\kappa\Omega_1\mathrm{e}^{-1/(\lambda_1-\mu_1^{*})}$, where \begin{eqnarray}\label{eq:mu1*}
\mu_1^{*}=(\mu^*-\lambda_2)/[1+(\mu^*-\lambda_2)\ln(\Omega_2/\Omega_1)]
\end{eqnarray}
is the ``pseudopotential'' defined at $\Omega_1$ and contributed by $(\mu^*-\lambda_2)$ defined at $\Omega_2$. (Here, we slightly generalize the concept of the pseudopotential to include both instantaneous and retarded interactions above a given frequency, which can also be understood as the ladder approximation in the pairing channel.)

\begin{figure}\begin{center}
  \includegraphics[width=1\textwidth]{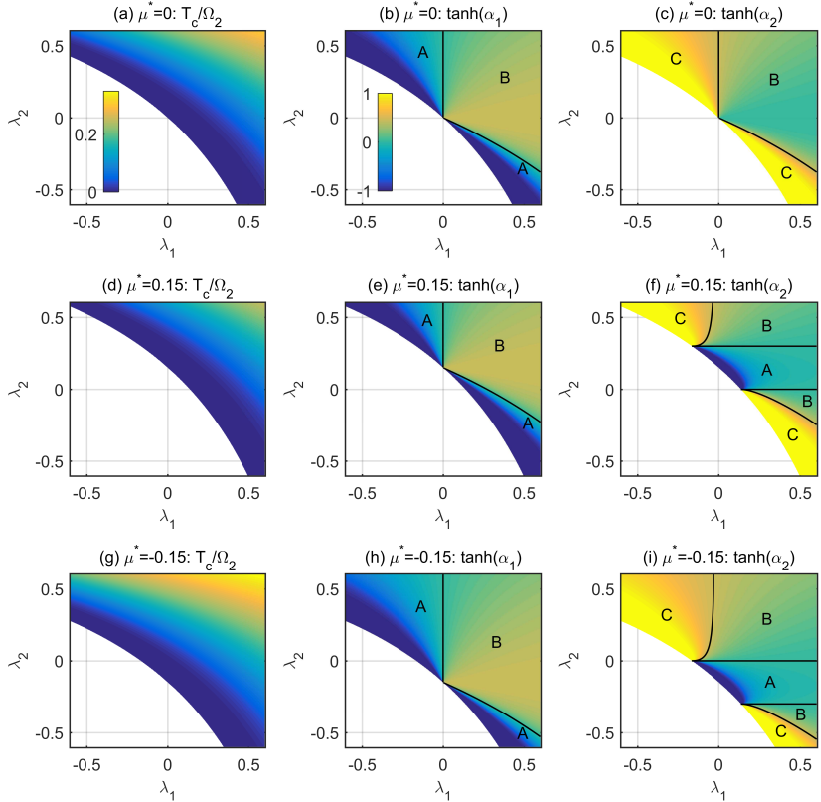}
  \caption{\label{fig:lam1_lam2} $T_c$ and $\alpha_{1,2}$ as functions of $\lambda_1$ and $\lambda_2$ by setting $\Omega_2=2\Omega_1$ and $\mu^*=0,0.15,-0.15$, respectively. For clarity, we divide the phase diagrams into different regimes: $\alpha<0$, $0<\alpha<1/2$ and $\alpha>1/2$ which are denoted as A, B and C, respectively. Since $\alpha_{1,2}$ ranges from $-\infty$ to $\infty$, we plot $\tanh(\alpha_{1,2})$ instead.}
\end{center}\end{figure}

Next, we turn to the isotope effect.
From Eq.~\ref{eq:Tc}, the isotope coefficient $\alpha_1$ or $\alpha_2$ can be obtained exactly, corresponding to $\Omega_1$ or $\Omega_2$ as the phonon mode, respectively:
\begin{eqnarray} \label{eq:alpha1}
\alpha_1=\frac12-\frac{(\mu^*-\lambda_2)^2}{2\left[\lambda_1+\lambda_2-\mu^*+\lambda_1(\mu^*-\lambda_2)\ln\left(\frac{\Omega_2}{\Omega_1}\right)\right]^2},
\end{eqnarray}
and
\begin{eqnarray} \label{eq:alpha2}
\alpha_2=\frac{\lambda_2(\lambda_2-2\mu^*)}{2\left[ \lambda_1+\lambda_2-\mu^*+\lambda_1(\mu^*-\lambda_2)\ln\left(\frac{\Omega_2}{\Omega_1}\right) \right]^2}.
\end{eqnarray}
Clearly, if both $\Omega_1$ and $\Omega_2$ are phonons, the total isotope coefficient $\alpha_1+\alpha_2$ is always less than $1/2$ in agreement with the Eliashberg theory. \cite{Carbotte1990}
Now, let us suppose only one of them is phonon.
In Fig.~\ref{fig:lam1_lam2}, we plot $T_c$ and $\alpha_{1,2}$ as functions of $\lambda_1$ and $\lambda_2$ by fixing $\Omega_2=2\Omega_1$ and $\mu^*=0,\pm0.15$, respectively.
Here, $\mu^*<0$ is also considered in order to describe instantaneous interaction induced superconductivity (\eg negative-$U$ Hubbard model).
For clarity, we divide the phase diagrams into several regimes: $\alpha_{1,2}<0$ (regime-A), $0<\alpha_{1,2}<1/2$ (regime-B), and $\alpha_{1,2}>1/2$ (regime-C).

Let's firstly examine $\alpha_1$ if $\Omega_1$ is the phonon, as shown in Figs.~\ref{fig:lam1_lam2}(b), (e) and (h).
Only regimes A and B appear, corresponding to $\alpha_1\le1/2$, which can be seen directly in its expression Eq.~\ref{eq:alpha1}.
An interesting feature is that as $T_c$ decreases $\alpha_1$ drops to negative values and finally diverges logarithmically: $\alpha_1\sim-\ln^2(T_c)$, unless in the critical point ($\lambda_1=0$ and $\lambda_2=\mu^*$).
This feature is already captured by the standard Eliashberg theory\cite{Carbotte1990}  when $\lambda_1>0$, as a result of the poorer screening of $\mu_1^{*}$ caused by increasing $\Omega_1$.
On the other hand, if $\lambda_1<0$, the negative $\alpha_1$ is expected as a result of its pair-breaking effect directly.

\begin{figure}\begin{center}
  \includegraphics[width=0.6\textwidth]{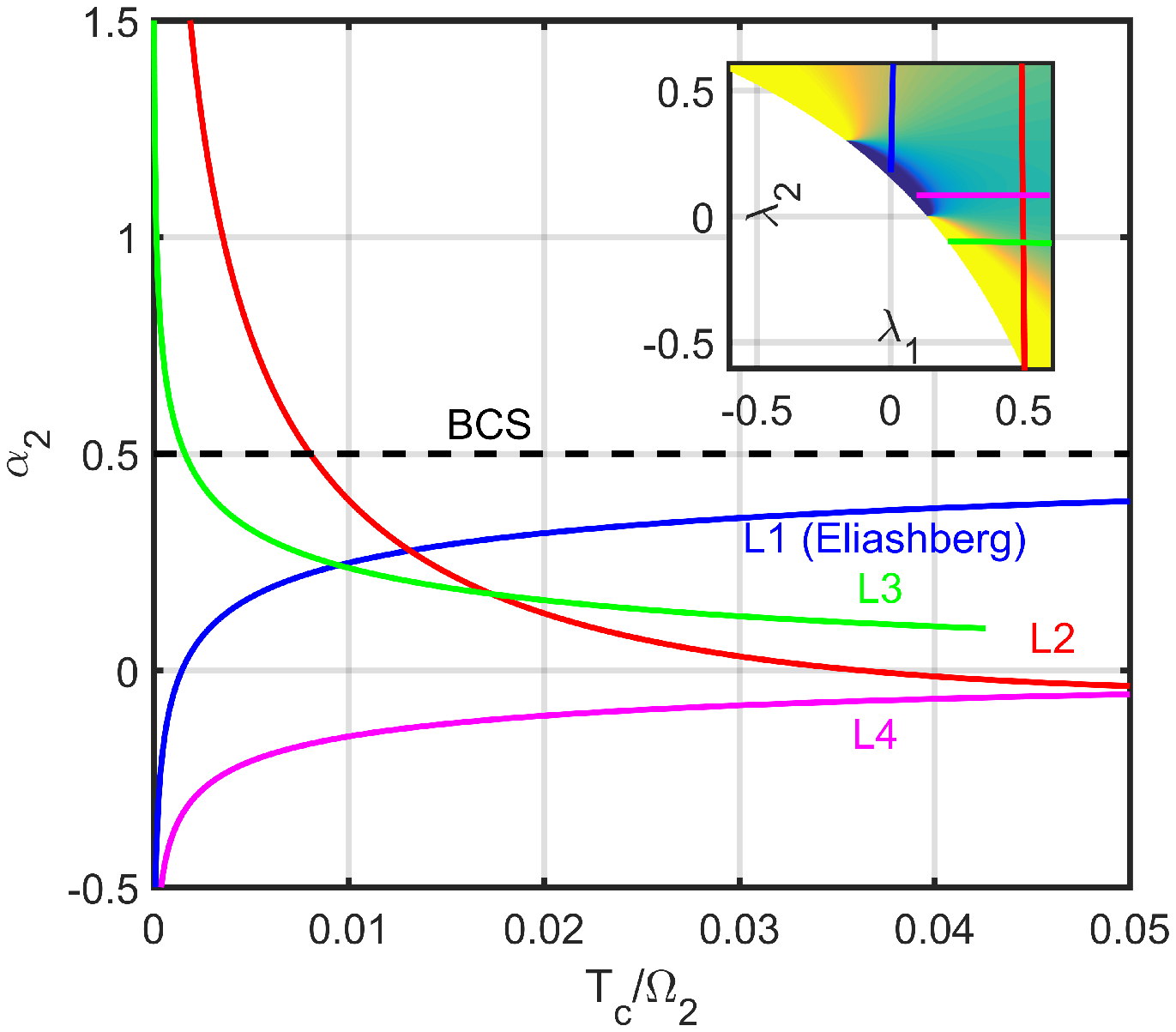}
  \caption{\label{fig:linecut} Typical scalings of $\alpha_2$ versus $T_c$ for several line cuts labeled by different colors as indicated in the inset, which is taken from Fig.~\ref{fig:lam1_lam2}(f).}
\end{center}\end{figure}

Our new result is in $\alpha_2$ if the higher frequency boson mode $\Omega_2$ is the phonon, as shown in Figs.~\ref{fig:lam1_lam2}(c), (f) and (i).
All three regimes can be found for nonzero $\mu^*$.
Large $\alpha_2>1/2$ (regime-C) is a ubiquitous feature for $\lambda_1\lambda_2<0$.
Sandwiched between them is the regime of normal isotope coefficient $0<\alpha_2<1/2$ (regime-B).
For $\mu^*\ne0$, there is also a regime of $\alpha_2<0$ (regime-A).
In fact, as seen from Eq.~\ref{eq:alpha2}, $\alpha_2\sim \lambda_2(\lambda_2-2\mu^*)\ln^2(T_c)$ as $T_c\rightarrow0$.
When $\lambda_2>0$, the large $\alpha_2$ originates from the pair-breaking effect of the $\Omega_1$ mode, as being discussed in Refs.~\cite{Carbotte1991,Dolgov2005,Harshman2008}.
Astonishingly, we have found another regime having large $\alpha_2$ for $\lambda_2<0$.
At first glance, this seems to be impossible since in this case the phonon is harmful to superconductivity.
How can a ``repulsive'' or pair-breaking phonon cause a large positive isotope effect?
The answer is: {\it the ``pseudopotential'' $\mu_1^{*}$ (Eq.~\ref{eq:mu1*}) contributed by $(\mu^*-\lambda_2)$ can be reduced by increasing $\Omega_2$, hence, leading to higher $T_c$.}
In Fig.~\ref{fig:linecut}, we plot several typical scalings of $\alpha_2$ versus $T_c$ along different paths as shown in the inset.
Line cut L1 is described by the standard Eliashberg theory by setting $\lambda_1=0$.
Both L2 and L3 give large $\alpha_2$ as $T_c$ decreases, while L2 also have negative $\alpha_2<0$ in the high $T_c$ regime.
Along L4, $\alpha_2$ is always negative and greatly enlarges the negative isotope coefficient regime of the standard Eliashberg theory (L1).
Discussions of these theoretical results in light of real materials are left to the next section.

In the above, we have found a universal scaling
\begin{eqnarray}
\alpha_{1,2}\rightarrow\pm\ln^2(T_c)
\end{eqnarray}
as $T_c\rightarrow0$ by tuning $\lambda_{1,2}$ to suppress the superconductivity.
In practice, there is another theoretical possibility to get $T_c\rightarrow0$ by tuning $\Omega_1\rightarrow0$ when $\lambda_1>0$ and $\lambda_2-\mu^*<0$.
In this case, we get the standard BCS result $\alpha_1=1/2$ and $\alpha_2\sim \ln^{-2}(\Omega_1)\rightarrow0$.

From the above discussions, we have seen that Eqs.~\ref{eq:alpha1} and \ref{eq:alpha2} give a full description of all possible values of the isotope coefficient in a BCS superconductor with two boson modes.
There are mainly three approximations in the above theory:
(1) We treat the boson mediated retarded interactions in momentum space directly. The frequency dependence of the gap functions is changed into momentum dependence effectively. This is just the standard BCS approximation. \cite{Bardeen1957}
(2) We have ignored band renormalization effect caused by the boson modes.
(3) We have assumed the pairing interactions and gap functions to be approximated by the three-piecewise functions.
In order to justify our approximations, we have performed numerical studies of the Eliashberg theory and obtained similar results, indicating the above BCS picture indeed works and captures the main physics qualitatively. See the appendix for more details.
In particular, the band renormalization effect can be approximately included in our BCS treatment.
The only difference is the kernel now becomes
\begin{eqnarray} \label{eq:K'}
K'=\left[\begin{array}{ccc}
Z_1^{-1} & 0 & 0 \\ 0 & Z_2^{-1} & 0 \\ 0 & 0 & 1 \end{array}
\right]K,
\end{eqnarray}
where $Z_1\approx1+\lambda_{1z}+\lambda_{2z}$ and $Z_2\approx1+\lambda_{2z}$.
It should be emphasized that for phonons $\lambda_z=\lambda_s$ and for magnetic modes $\lambda_z=-\lambda_s$ where $\lambda_s$ denotes the electron-boson coupling constant {\bf in the uniform s-wave paring channel}. (See the appendix for more details.)
We have checked that including the band renormalization effect does not change the above BCS results qualitatively.
The derivations of $T_c$ and $\alpha_{1,2}$ are straightforward and thus left out. Their analytical expressions can be found in the appendix.
Another approximation in our study is that both boson modes are taken as Einstein modes, i.e., single-energy modes.
For continuous boson modes, $\Omega_1$ and $\Omega_2$ can be understood as their representative energies.
In most materials, the phonon and non-phonon modes range in different energy regions, and hence, we expect the above Einstein mode approximations are qualitatively correct.

\section{Summary and discussions}
In summary, we have found the anomalous isotope effect ($\alpha<0$ or $\alpha>1/2$) can be explained in the BCS theory when both phonon and non-phonon modes coexist.
If the phonon frequency is lower, $\alpha\leq1/2$. But if the phonon has a higher frequency, any values of $\alpha$ can be obtained.
Interestingly, we have obtained a scaling behavior $\alpha\rightarrow\pm\ln^2(T_c)$ as $T_c\rightarrow0$ by tuning $\lambda_{1,2}$ to suppress the superconductivity.
Most strikingly, when the phonon mode has a higher frequency, $\alpha$ can be larger than $1/2$ even if it is pair-breaking.

Finally, as possible applications, we provide some remarks about several superconducting materials from the viewpoint of the BCS framework.
(1) Cuprates.
Low energy boson modes have been widely observed in many different experiments in cuprates, including mainly two candidates: phonon and magnetic modes. \cite{Zhou2007,Tranquada2007,Carbotte2011}
We have listed several candidate boson modes in the appendix.
Taking different experiments together, roughly speaking, the 70 meV kink observed in angle-resolved photoemission spectral (ARPES) can be assigned to the breathing phonon \cite{Lanzara2001,Wang2002,Iwasawa2008} and the low energy (10meV$\sim$60meV including the famous 41meV resonance\cite{Rossat-Mignod1991}) \cite{Tranquada2007,Fujita2012,Carbotte2011} are dominated by antiferromagnetic (AF) excitations with the hour glass dispersion \cite{Tranquada2004}, although phonon may also have some contributions. \cite{Cuk2004,Devereaux2004,Gweon2004,Lee2006,Douglas2007,He2018}
Since the breathing phonon is against d-wave superconductivity \cite{Bulut1996} but the AF fluctuation can mediate d-wave superconductivity \cite{Scalapino1986,Monthoux1991,Moriya2000,Scalapino2012}, within our BCS picture, they correspond to $10$meV$<\Omega_1<60$meV, $\Omega_2\sim70$meV,  $\lambda_1>0$ and $\lambda_2<0$.
{For a rough but specific estimation, if we choose $\Omega_1\sim35$meV, $\Omega_2\sim70$meV, $\lambda_2\sim-0.1$, then the maximal $T_c\sim80$K at optimal doping corresponds to $\lambda_1\sim0.8$.
Upon doping away from the optimal doping, $\lambda_1$ drops to reduce $T_c$ and enhance $\alpha_2$ and finally leads to the scaling behavior $\alpha_2\sim\ln^2(T_c)$ as $T_c\rightarrow0$.}
This behavior is similar to the experimental observations. \cite{Keller2005,Keller2008}
In our theory, adding $\mu^*$ does not change the qualitative behavior and thus is not in contradictory with the resonating valence bond (RVB) theory \cite{Anderson1987} which corresponds to an attractive pseudopotential $\mu^*<0$.
But based on our picture, the lower energy boson mode is necessary to obtain $\alpha_2>1/2$.
In particular, for La$_{2-x}$Sr$_x$CuO$_4$ near 1/8 doping\cite{Crawford1990} compared with YBa$_2$Cu$_3$O$_y$\cite{Kamiya2014}, stronger charge fluctuation results in weaker $\lambda_1$ and leads to smaller $T_c$ and larger $\alpha_2$, also in agreement with the experiment.
(2) Sr$_2$RuO$_4$.
Ferromagnetic (FM) fluctuations are widely believed to mediate the superconductivity in Sr$_2$RuO$_4$. \cite{Mackenzie2003}
The magnetic mode energy is found to be less than $15$meV, \cite{Sidis1999} much less than the O-phonon frequency around $50$meV. \cite{Mao2001}
Therefore, it is in the same parameter regime as cuprates.
As a result, its $\alpha$ versus $T_c$ shows similar behavior as cuprates except $\alpha_2<0$ for higher $T_c$ samples \cite{Mao2001}.
(3) Iron-based superconductors are found to be similar to cuprates in the sense that AF fluctuations are closely related to superconductivity (either s$_\pm$- or d-wave). \cite{Stewart2011,Chubukov2012}
If we take the AF fluctuation as $\Omega_1\sim15$meV \cite{Christianson2008,Dai2015} and phonon as $\Omega_2\sim40$meV \cite{Liu2009}, we can obtain both $\alpha_2<0$ \cite{Shirage2009} or $\alpha_2>1/2$ \cite{Khasanov2010}.
However, further systematic experiments of the isotope effect upon doping are needed to pin down the role of phonons.
(4) C$_{60}$-based superconductors.
In fullerides superconductors A$_3$C$_{60}$ (A stands for K, Rb, Cs), phonon mediated s-wave pairing is widely accepted. \cite{Gunnarsson1997} $\alpha>1/2$ has been reported \cite{GinsbergBook,Ricco2008} and explained by the breakdown of Migdal theorem. \cite{Grimaldi1995,Grimaldi1995a} Nevertheless, our theory provides another possibility: existence of a lower frequency non-phonon mode can also lead to large $\alpha$. Interestingly, in A15-Cs$_3$C$_{60}$ superconductivity is found to be near the AF parent \cite{Takabayashi2009} such that the spin fluctuation may also play some role in it. \cite{Nomura2015}

Although the above discussions are based on the BCS framework, we hope the qualitative conclusions should be the leading effect even beyond the weak-coupling limit.
Therefore, we emphasize that the isotope effect can be strongly affected by the non-phonon boson mode.

\ack
D.W. thanks Shun-Li Yu for discussions on spin fluctuations in cuprates.
This work is supported by National Natural Science Foundation of China (under Grant Nos. 11874205 and 11574134) and National Key Research and Development Program of China (under Grant No. 2016YFA0300401).

\appendix
\section{Eliashberg theory}
In this section, we make a benchmark for our BCS approach by numerically solving the Eliashberg equations for general pairing symmetries. At first, for simplicity, we consider only one phonon mode and give a self-contained derivation. The Eliashberg theory is based on the self energy 
in Nambu space, \cite{Nambu1960,Scalapino1966,Carbotte1990}
\begin{eqnarray}\label{eq:selfconsist}
\Sigma(p)=-\frac{T}{N}\sum_{p'} \sigma_3 G(p')\sigma_3 D(p-p')|g(p-p')|^2,
\end{eqnarray}
where $G$/$D$ are the electron/phonon propagators, $g$ is the electron-phonon vertex, and $p$($p'$) stands for both momentum and frequency. $\sigma_3$ is the third Pauli matrix.
By choosing the ansatz:
\begin{eqnarray}
\Sigma(p)=(1-Z)i\omega_n\sigma_0+{\rm Re}(\phi)\sigma_1+{\rm Im}(\phi)\sigma_2,
\end{eqnarray}
and comparing two sides of Eq.~\ref{eq:selfconsist}, we get (particle-hole symmetry is assumed here and can be generalized straightforwardly)
\begin{eqnarray}
[Z(p)-1]i\omega_n &=-\frac{T}{N}\sum_{p'} |g(p-p')|^2D(p-p')\frac{Z(p')i\omega_n'}{Z(p')^2\omega_n'^2+\varepsilon_{\0p'}^2+|\phi(p')|^2}, \label{eq:self-Z}\\
\phi(p) &= -\frac{T}{N}\sum_{p'} |g(p-p')|^2D(p-p')\frac{\phi(p')}{Z(p')^2\omega_n'^2+\varepsilon_{\0p'}^2+|\phi(p')|^2} \label{eq:self-phi}.
\end{eqnarray}
Then, singular mode decomposition is performed for $|g|^2D$ such that $|g(\0p-\0p')|^2=\sum_m g_{m}^2f_m(\0p)f_m(\0p')$ where $f_m(\0p)$ are form factors in different symmetry channels. Next, we take two other ansatzs \cite{Scalapino1966}:
\begin{eqnarray}
\phi(\0p,i\omega_n)&=f_m(\0p)Z(i\omega_n)\Delta_m(i\omega_n), \\ Z(\0p,i\omega_n)&=Z(i\omega_n).
\end{eqnarray}
These assumptions are justified by the facts: (1) non s-wave (momentum independent) component of $Z$ is small due to the momentum summation of its self-consistent equation \ref{eq:self-Z}. (2) the gap function $\Delta=\phi/Z$ is determined only by the uniform s-wave component of $Z$ up to the leading order.
Completing the momentum summations of Eqs.~\ref{eq:self-Z} and \ref{eq:self-phi} by energy integration with constant density of states, we obtain
\begin{eqnarray}
[Z(i\omega_n)-1]i\omega_n &=\pi T\sum_{i\omega_n'}\frac{\lambda_z(i\omega_n-i\omega_n')i\omega_n'}{\sqrt{\omega_n'^2+|\Delta_m(i\omega_n')|^2}}, \\
Z(i\omega_n)\Delta_m(i\omega_n) &=\pi T\int\frac{\mathrm{d}\Omega}{4\pi^2}\sum_{i\omega_n'}\frac{\lambda_m(i\omega_n-i\omega_n')f_m^2(\Omega)\Delta_m(i\omega_n')}{\sqrt{\omega_n'^2+f_m(\Omega)^2|\Delta_m(i\omega_n')|^2}} ,
\end{eqnarray}
where $\Omega$ is the solid angle {(not confused with Debye frequency $\Omega_D$ or $\Omega_{1,2}$)} and
\begin{eqnarray}
\lambda_m(i\omega_n-i\omega_n')=-\+Ng_m^2 D(i\omega_n-i\omega_n'),
\end{eqnarray}
and $\lambda_z$ is the uniform s-wave component.
{\bf Notice that only uniform s-wave component of $\lambda_s$ enters the self-consistent equation of $Z$ while $\lambda_m$ enters into the gap self-consistent equation. This is a fundamental difference between unconventional and conventional superconductors.}
The frequency summation should be bounded by the Fermi energy $E_F$ as a result of the factor $\arctan(E_F/|\omega_n'|)$ (not shown explicitly).

At $T=T_c^-$, $\Delta_m\rightarrow0$, we can absorb the phase factor $f_m(\Omega)$ into $\tilde{\lambda}_m=\lambda_m\av{f_m^2(\Omega)}$.
Then, the self-consistent equations are linearized as
\begin{eqnarray}
[Z(i\omega_n)-1]\omega_n &=\pi T_c\sum_{i\omega_n'}\lambda_z(i\omega_n-i\omega_n')\mathrm{sgn}(\omega_n'), \label{eq:linear-Z}\\
Z(i\omega_n)\Delta_m(i\omega_n)&=\pi T_c\sum_{i\omega_n'}\frac{\tilde{\lambda}_m(i\omega_n-i\omega_n')\Delta_m(i\omega_n')}{|\omega_n'|} .\label{eq:linear-Delta}
\end{eqnarray}
In the following, we neglect the tilde symbol in $\tilde{\lambda}_m$ for simplicity.

For magnetic modes, the vertex $\sigma_3$ in Eq.~\ref{eq:selfconsist} should be replaced by $\sigma_0$. Then, there will be an additional minus sign in the right hand side of Eq.~\ref{eq:linear-Delta}. {\bf For magnetic boson, we absorb the minus sign in the definition of $\lambda_m$ for all $m$ and keeps Eqs.~\ref{eq:linear-Z} and \ref{eq:linear-Delta} unchanged. But the price is $\lambda_z=-\lambda_s$ for magnetic modes}.

Before going on, a short discussion on the Coulomb pseudopotential is given. Coulomb pseudopotential $\mu$ is not others but a boson mode with infinite frequency such that its $\lambda_m$ is frequency independent. Therefore, it has no contribution to $Z$ but has to be included in the self-consistent equation of $\Delta_m$. In practice, the Coulomb pseudopotential may also be defined at a middle frequency $\omega_c$ satisfying $\Omega_D\ll\omega_c\ll E_F$ with $\mu(\omega_c)=\mu(E_F)/[1+\mu(E_F)\log(E_F/\omega_c)]$. \cite{Morel1962,McMillan1968,Carbotte1990}

\begin{figure}\begin{center}
  \includegraphics[width=0.3\textwidth]{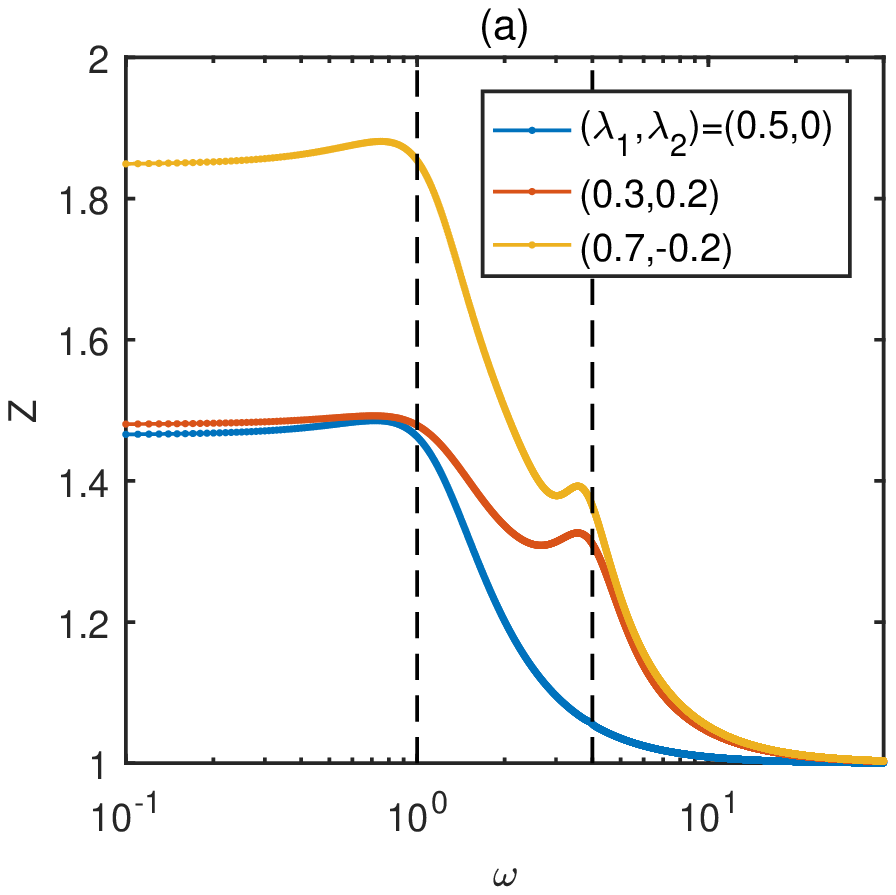}
  \includegraphics[width=0.3\textwidth]{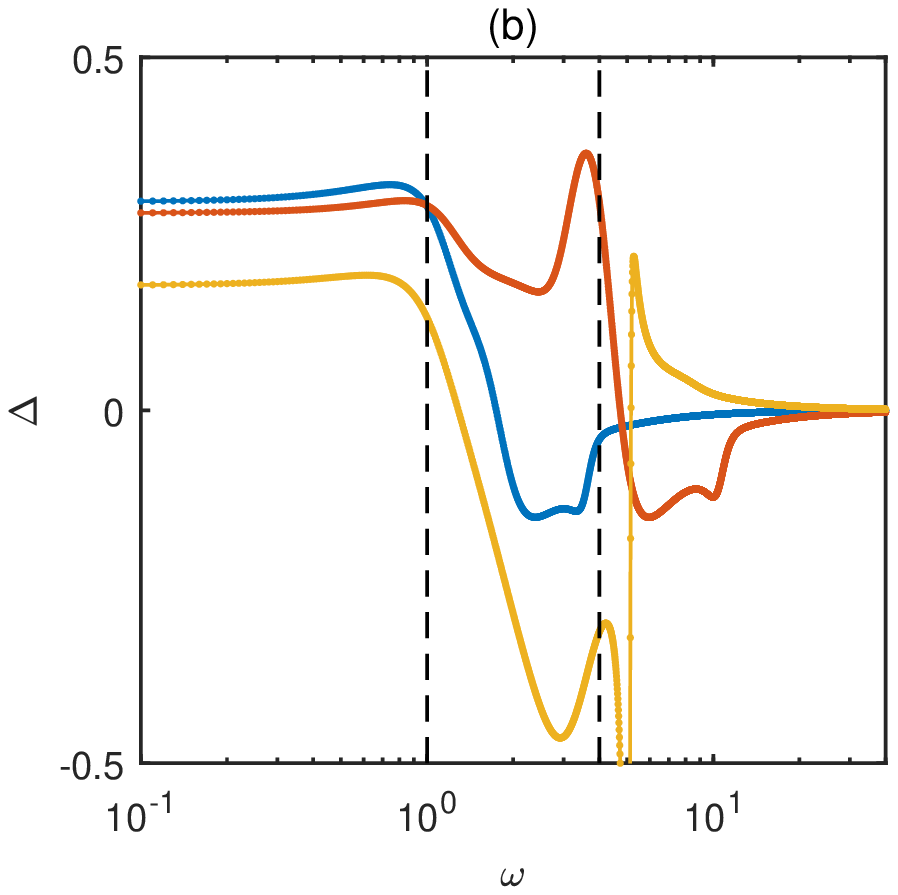}
  \includegraphics[width=0.3\textwidth]{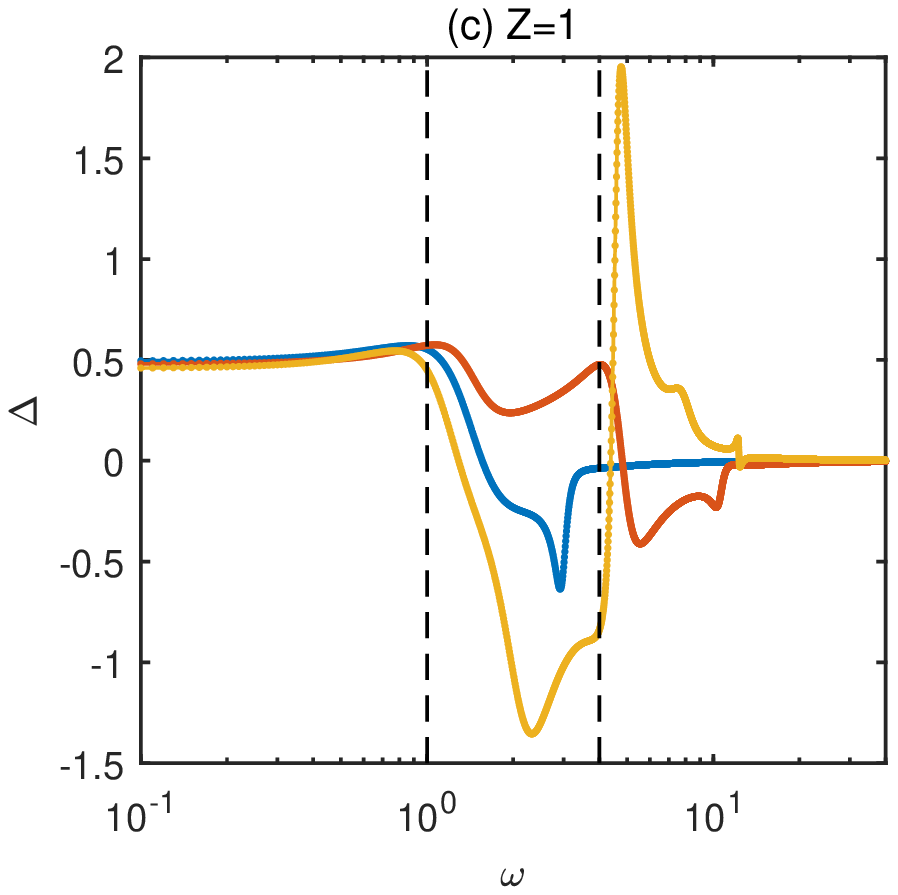}
  \caption{\label{fig:check3well} Results of $Z(\omega)$ and $\Delta(\omega)$ by solving imaginary frequency Eliashberg equations and analytical continuation using the Pad\'e approximation in the case of $\mu=0$. (a) and (b) are obtained by setting $\lambda_{iz}=|\lambda_i|$. (c) is for $\lambda_{iz}=0$, \ie ignoring band renormalization. Dashed lines indicate two boson modes $\Omega_2=4\Omega_1$.}
\end{center}\end{figure}

\begin{figure}\begin{center}
  \includegraphics[width=0.8\textwidth]{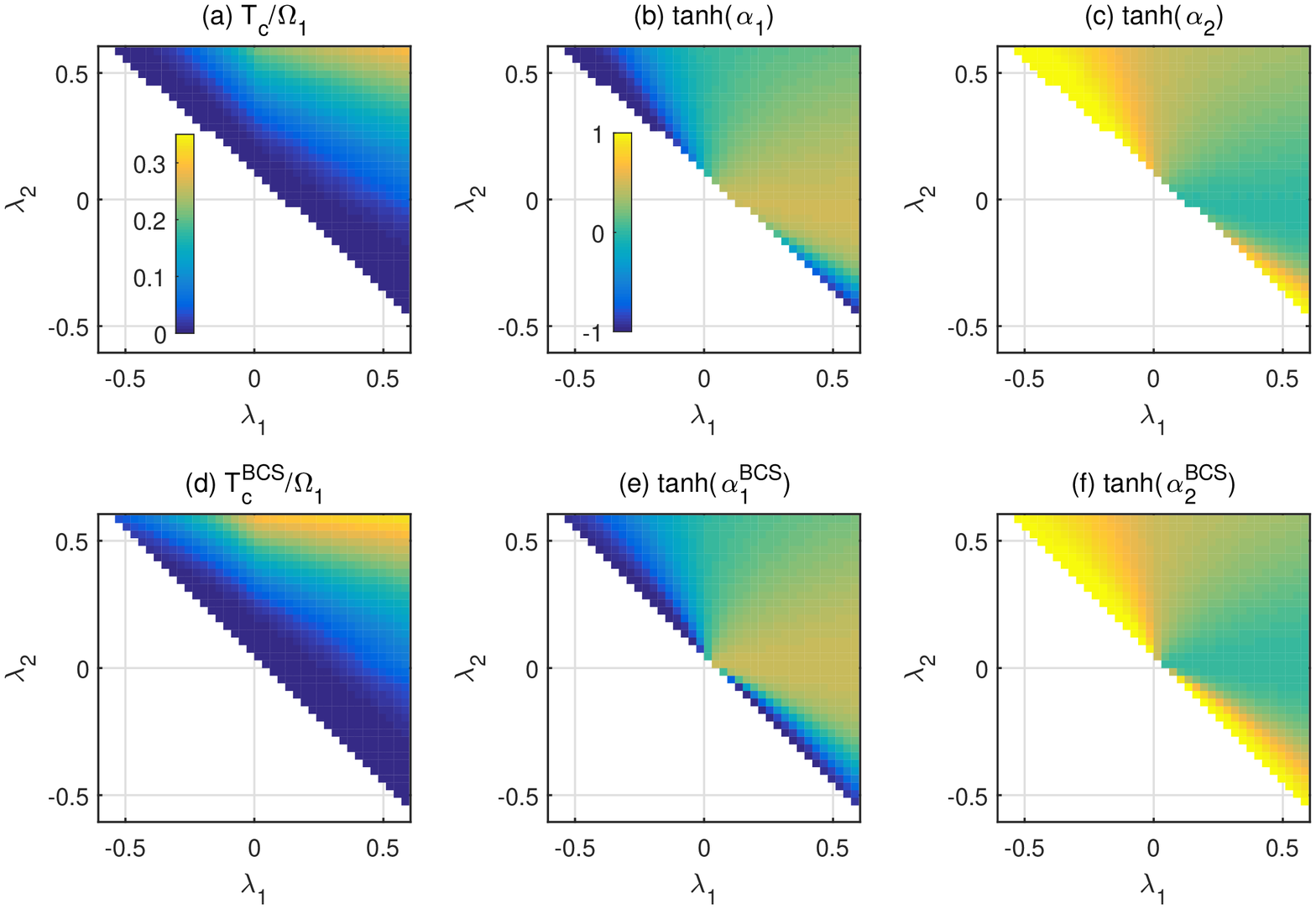}
  \caption{\label{fig:elia_vs_bcs} Phase diagrams obtained by Eliashberg equations in (a-c), and BCS theory in (d-f). In the calculations, $\Omega_2=4\Omega_1$ and $\mu^*=0$ are used. Notice that for the Eliashberg calculations, in practice, the values of $T_c$ cannot be made arbitrarily small due to the number of Matsubara frequencies cannot be chosen arbitrarily large.}
\end{center}\end{figure}

In practice, Eq.~\ref{eq:linear-Z} is firstly solved to obtain $Z(\omega_n)$ numerically. Then, $T_c$ is obtained by finding the largest eigenvalue of the kernel
\begin{eqnarray}
K(\omega_n,\omega_n')=\frac{\pi T_c}{Z(i\omega_n)}\sum_{i\omega_n'}\frac{\tilde{\lambda}_m(i\omega_n-i\omega_n')}{|\omega_n'|}
\end{eqnarray}
to be $1$ and $\Delta_m$ is given by the eigenvector.
After obtaining imaginary frequency data $Z(i\omega_n)$ and $\Delta(i\omega_n)$, we perform the analytical continuation using the Pad\'e approximation. \cite{Vidberg1977}
The results are shown in Fig.~\ref{fig:check3well} by setting $\lambda_{iz}=|\lambda_{i}|$ in (b) and $\lambda_{iz}=0$ in (c), respectively.
$Z(\omega)$ and $\Delta(\omega)$ are found to show drastic change near $\Omega_1$ and $\Omega_2$, supporting our three-piecewise approximation in the main text.
As a further benchmark, we also present the phase diagrams on the $\lambda_1-\lambda_2$ plane in Fig.~\ref{fig:elia_vs_bcs}, which are in quite good agreement with the BCS theory.

\section{Modified BCS theory}
In this section, we show the results of the modified BCS theory by considering the band renormalization effect. In this case, the kernel is given by Eq.~\ref{eq:K'} in the main text. Following the method in the main text, $T_c$ and $\alpha_{1,2}$ can be obtained as follows
\begin{eqnarray}
T_c=\kappa\Omega_1\exp\left\{ -\frac{Z_1\left[Z_2+(\mu^*-\lambda_2)\ln\left(\frac{\Omega_2}{\Omega_1}\right)\right]}{Z_2(\lambda_1+\lambda_2-\mu^*)+\lambda_1(\mu^*-\lambda_2)\ln\left(\frac{\Omega_2}{\Omega_1}\right)} \right\},
\end{eqnarray}
and
\begin{eqnarray}
\alpha_1=\frac12-\frac{Z_1Z_2(\mu^*-\lambda_2)^2}{2\left[Z_2(\lambda_1+\lambda_2-\mu^*)+\lambda_1(\mu^*-\lambda_2)\ln\left(\frac{\Omega_2}{\Omega_1}\right)\right]^2},
\end{eqnarray}
and
\begin{eqnarray}
\alpha_2=\frac{Z_1Z_2\left[\lambda_2(\lambda_2-2\mu^*)+\mu^{*2}(Z_2-1)\right]}{2\left[Z_2( \lambda_1+\lambda_2-\mu^*)+\lambda_1(\mu^*-\lambda_2)\ln\left(\frac{\Omega_2}{\Omega_1}\right) \right]^2}.
\end{eqnarray}
Notice that the scaling behavior of $\alpha_{1,2}\sim\pm\ln^2(T_c)$ as $T_c\rightarrow0$ keeps unchanged.

\section{Several boson modes in cuprates}
In this section, we list several typical phonon and magnetic modes in cuprates in table \ref{tb:lam} together with three instantaneous interactions which can be taken as the pseudopotentials. All phonon modes have positive $\lambda_s$ and all magnetic modes have negative $\lambda_s$. Therefore, for conventional uniform s-wave superconductors, phonon can mediate superconductivity but the magnetic modes only cause pair-breaking. Quite differently, for unconventional superconductors, both phonon and magnetic modes can be either positive or negative depending on different pairing symmetries.

\begin{table} \caption{\label{tb:lam} Electron boson couplings of several typical phonon (Holstein, breathing and buckling) and magnetic (AF and FM fluctuations) boson modes in the superconducting channel in cuprates. Instantaneous interactions (Hubbard, Heisenberg, and nearest neighbour Coulomb) as the Coulomb pseudopotentials are also listed below for which $\lambda_m=-\mu_m$.}
  \begin{center}
    \begin{tabular}{|c|c|c|}
      \hline
      & $\lambda_s$ & $\lambda_d$ \\ \hline
      Holstein       &      +      &      0      \\
      breathing      &      +      &     $-$     \\
      B$_{1g}$-buckling     &      +      &      +      \\
      AF fluctuation    &     $-$     &      +      \\
      FM fluctuation    &     $-$     &     $-$     \\
      $Un_{i\up}n_{i\dn}$ &     $-$     &      0      \\
      $J\0S_i\cdot\0S_j$  &      +      &      +      \\
      $Vn_in_j$      &     $-$     &     $-$     \\ \hline
    \end{tabular}
  \end{center}
\end{table}

For the d$_{x^2-y^2}$-wave pairing in cuprates, B$_{1g}$-buckling phonon mode has a positive $\lambda_d$ due to its form factor $\left[\cos^2(q_x/2)+\cos^2(q_y/2)\right]$ and has been used as one candidate of the pairing mechanism. \cite{Bulut1996,Devereaux2004,Honerkamp2007,Johnston2010} However, the buckling mode requires the mirror symmetry breaking \cite{Bulut1996} and does not exist in single layer cuprates.
Differently, the breathing phonon mode always exists and has been evidenced in ARPES experiments as the 70meV kink, \cite{Lanzara2001,Iwasawa2008} which is in fact against d-wave SC since its $\lambda_d<0$ as a result of its form factor $\left[\sin^2(q_x/2)+\sin^2(q_y/2)\right]$. \cite{Bulut1996,Johnston2010}
Besides, the Holstein phonon has no direct d-wave component and thus can be neglected in the leading order approximation without considering its coupling to other interaction channels.

On the other hand, the magnetic fluctuations are widely observed \cite{Tranquada2007,Fujita2012,Carbotte2011} and believed to be closely related to the d-wave superconductivity, including the spin fluctuation mechanism \cite{Scalapino1986,Monthoux1991,Moriya2000,Scalapino2012} and the emergent effective SO(5) symmetry of the t-J model. \cite{Zhang1997,Demler2004} Combining most experiments, especially neutron and ARPES, it's reasonable to assume its energy ranging from 10meV to 60meV. \cite{Tranquada2007,Zhou2007,Carbotte2011,Fujita2012} For simplicity, we have considered two extreme cases: AF and FM fluctuations with opposite $\lambda_d$.

Finally, in order to include the instantaneous interactions, we also consider three interaction terms: Hubbard, Heisenberg exchange and Coulomb interaction between nearest neighboring sites. Although the Hubbard term has no d-wave pairing interaction directly, the Heisenberg term does have attractive component in the d-wave pairing channel, which in fact plays the essential role of pairing in the RVB theory. \cite{Anderson1987,Baskaran1987,Zhang1988,Lee2006Mott} In addition, the nearest neighbour Coulomb interaction (which already exists in the t-J model) leads to a repulsive pairing interaction and thus should contribute to a positive Coulomb pseudopotential.


\section*{References}
\bibliographystyle{iopart-num}
\bibliography{isotope}

\end{document}